\documentclass[letter]{emulateapj}
\usepackage{apjfonts}
\usepackage{graphicx}
\usepackage{natbib}
\usepackage{float}
\usepackage{color}

\begin{document}
  \title{Normal Type Ia supernovae from violent mergers of white dwarf binaries}
  
  \author{R.~Pakmor}
  \affil{Heidelberger Institut f\"{u}r Theoretische Studien, Schloss-Wolfsbrunnenweg 35, 69118 Heidelberg, Germany}
  
  \author{M.~Kromer, S.~Taubenberger}
  \affil{Max-Planck-Institut f\"{u}r Astrophysik, Karl-Schwarzschild-Str. 1, 85741 Garching, Germany}
  
  \author{S.~A.~Sim}
  \affil{Research School of Astronomy and Astrophysics, Mount Stromlo Observatory, Cotter Road, Weston Creek, ACT 2611, Australia}
   
  \author{F.~K.~R\"{o}pke}
  \affil{Universit{\"a}t W{\"u}rzburg, Emil-Fischer-Str. 31, 97074 W{\"u}rzburg, Germany}
   
  \author{W.~Hillebrandt}
  \affil{Max-Planck-Institut f\"{u}r Astrophysik, Karl-Schwarzschild-Str. 1, 85741 Garching, Germany}

  \begin{abstract}
  
  One of the most important questions regarding the progenitor systems of Type Ia
  supernovae (SNe Ia) is whether mergers of two white dwarfs can lead to explosions
  that reproduce observations of normal events.   
  Here we present a fully three-dimensional simulation of a violent merger of two
  carbon-oxygen white dwarfs with masses of $0.9 \mathrm{M_\odot}$ and $1.1 \mathrm{M_\odot}$
  combining very high resolution and exact initial conditions.
  A well-tested combination of codes is used to study the system. We start with the
  dynamical inspiral phase and follow the subsequent thermonuclear explosion under the plausible
  assumption that a detonation forms in the process of merging. We then perform
  detailed nucleosynthesis calculations and radiative transfer simulations to predict synthetic
  observables from the homologously expanding supernova ejecta.
  We find that synthetic color lightcurves of our merger, which produces about
  $0.62  \mathrm{M_\odot}$ of $^{56}\mathrm{Ni}$, show good agreement with those
  observed for normal SNe Ia in all wave bands from U to K. Line velocities 
  in synthetic spectra around maximum light also agree well with observations.
  We conclude, that violent mergers of massive white dwarfs can closely resemble normal
  SNe Ia. Therefore, depending on the number of such massive systems available
  these mergers may contribute at least a small fraction to the observed population
  of normal SNe Ia.
  
  \end{abstract}


  \shortauthors{R.~Pakmor et al.}

  \maketitle

  \section{Introduction}
  \label{sec:introduction}
  
  Despite many years of dedicated research, the progenitor systems and explosion mechanisms of SNe Ia
  remain unclear. It is generally accepted that SNe Ia originate from thermonuclear explosions of massive
  carbon-oxygen white dwarfs in binary systems. Depending on the nature of the companion star, two different
  progenitor systems have been proposed. In the \emph{single degenerate} scenario \citep{whelan1973a} a
  carbon-oxygen white dwarf accretes from a non-degenerate companion star until it reaches the Chandrasekhar-mass
  and explodes \citep[but note that an explosion before reaching the Chandrasekhar mass may also be possible,
  e.g.][]{fink2007a,fink2010a}. In contrast, in the \emph{double degenerate} scenario \citep{iben1984a, webbink1984a} a
  merger of two carbon-oxygen white dwarfs causes a thermonuclear explosion of the merged system.
  
  Many recent findings including SN rates from population synthesis studies \citep{ruiter2009a}, studies of the delay
  time distribution of observed SNe Ia \citep[e.g.][]{maoz2010a}, the lack of radio emission of SNe Ia
  \citep[e.g.][]{hancock2011a}, the lack of hydrogen emission in nebular spectra of SNe Ia \citep{leonard2007a},
  the lack of X-ray emission in elliptical galaxies \citep{gilfanov2010a} and studies of SN Ia remnants
  \citep[e.g.][]{badenes2007a} seem to favor the double degenerate scenario. In addition, first studies of
  SN~2011fe seem to disfavor a single degenerate progenitor \citep[see, e.g.][]{bloom2011a}. 
  There is, however, no unambiguous proof for any progenitor scenario yet. For a detailed discussion about
  constraints on the progenitor scenarios see \citet{howell2011a}.
  
  To date, most theoretical work on SNe Ia has concentrated on the single degenerate scenario, as mergers of
  white dwarfs were thought not to lead to thermonuclear explosions. This was mainly based on the picture that
  mergers would leave behind the more massive white dwarf with a hot envelope made up of the material of the less 
  massive white dwarf. Fast accretion from the envelope will then turn the carbon-oxygen white dwarf into an
  oxygen-neon white dwarf \citep{nomoto1985a,saio1998a} which collapses to a neutron star as it approaches
  the Chandrasekhar mass \citep{nomoto1991a}. Ways to avoid the transformation of the carbon-oxygen white
  dwarf into an oxygen-neon white dwarf have been proposed \citep[e.g.][]{yoon2007a}, but without conclusive results.
  
  \begin{figure*}
     \centering
     \includegraphics[width=0.9\linewidth]{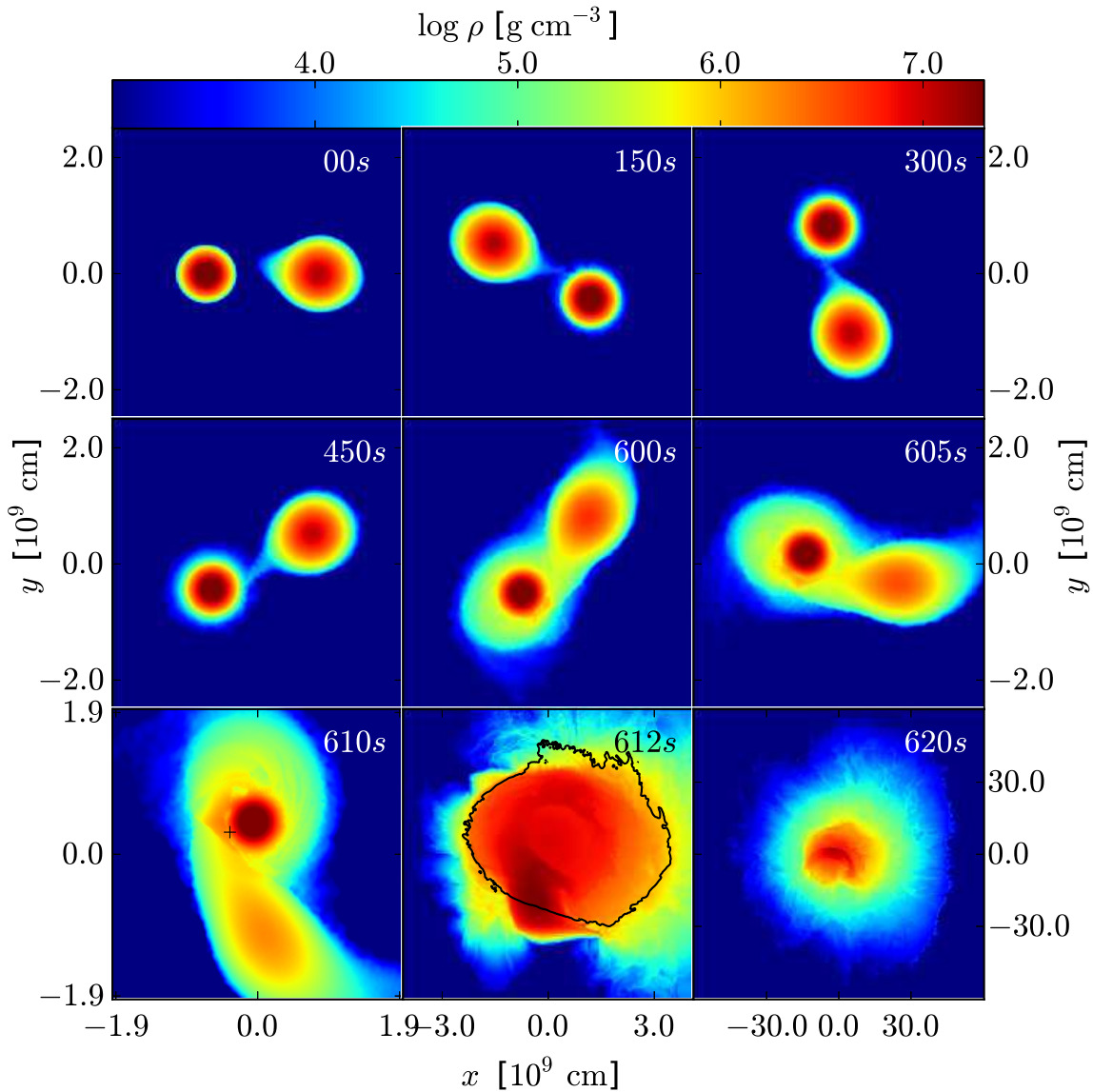}
     \caption{Snapshots of the merger of a $1.1\, \mathrm{M}_\odot$ and a $0.9\, \mathrm{M}_\odot$ carbon-oxygen
     white dwarf and the subsequent thermonuclear explosion. At the start of the simulation the binary system has an
     orbital period of  $\approx 35\, \mathrm{s}$. The black cross indicates the position where the detonation is ignited.
     The black line shows the position of the detonation front. Color-coded is the logarithm of the density. The last two
     panels have a different color scale ranging from $10^{-4}\,\mathrm{g\ cm^{-3}}$ to $10^{6}\,\mathrm{g\ cm^{-3}}$
     and $10^{4}\,\mathrm{g\ cm^{-3}}$, respectively.}
     \label{fig:merger}
  \end{figure*}
  
  Recently, however, we demonstrated that violent mergers of two carbon-oxygen white dwarfs could directly
  lead to a thermonuclear explosion while the merger is still ongoing \citep{pakmor2010a}. We also showed that
  the observables for such an explosion with two white dwarfs of $0.9\, \mathrm{M_\odot}$ show good
  agreement with the observed properties of subluminous 1991bg-like SNe Ia. Furthermore, we found that for a primary mass of 
  $0.9\, \mathrm{M_\odot}$ a mass ratio of at least about $0.8$ is required for the scenario to work \citep{pakmor2011b}.
  
  Lately, \citet{dan2011a} showed that using exact initial conditions can change the properties of the merger. In particular,
  this leads to a significantly longer inspiral in their simulations. However, \citet{dan2011a} were only able to run the simulation with
  a resolution of $2 \times 10^5$ particles (for comparison, the violent merger calculations by \citet{pakmor2010a,pakmor2011b}
  used $2 \times 10^6$ particles).
  
  In this work, we combine high resolution merger simulations with exact initial conditions.
  We present the results of a simulation of the massive merger of a $1.1\, \mathrm{M_\odot}$
  and a $0.9\, \mathrm{M_\odot}$ carbon-oxygen white dwarf. We follow the evolution of the binary system
  through the merger phase, thermonuclear explosion and nucleosynthesis.
  Finally we use three-dimensional radiative transfer simulations to obtain synthetic lightcurves and spectra.

  \section{Merger and Explosion}
  \label{sec:merger}
  
  The inspiral and merger is modeled using a modified version of the \textsc{gadget}
  code \citep{springel2005a}. Modifications include the Helmholtz equation
  of state \citep{timmes2000a} and a 13 isotope nuclear reaction network
  that contains all $\mathrm{\alpha}$-elements from $^{4}\mathrm{He}$ to $^{56}\mathrm{Ni}$.
  Radiative cooling effects are not included in our simulation.
  A detailed description of the modifications will be given in a forthcoming paper. In addition
  the maximum smoothing length of a particle was restricted to $10^8\,\mathrm{cm}$. This affects
  only particles ejected from the binary system during the merger but leads to a 
  significant speedup of the code. Since these particles are at very low densities and contain only
  less than one percent of the total mass, they have no noticable influence on the explosion dynamics and observables.
  
  The initial binary system consists of a $1.1\, \mathrm{M_\odot}$ and a $0.9\, \mathrm{M_\odot}$ carbon-oxygen
  white dwarf constructed from a total of $1.8 \times 10^6$ equal-mass particles. Both white dwarfs are set up in isolation and
  relaxed with an additional friction force for $100\, \mathrm{s}$. We then apply the method described in \citet{dan2011a}
  to slowly move the two white dwarfs close together. When the first particle of the less massive white dwarf crosses
  the inner Lagrange-point we stop and start the actual simulation. At this time, the binary system has an orbital period
  of about $35\, \mathrm{s}$.
  
  \begin{figure}
     \centering
     \includegraphics[width=0.9\linewidth]{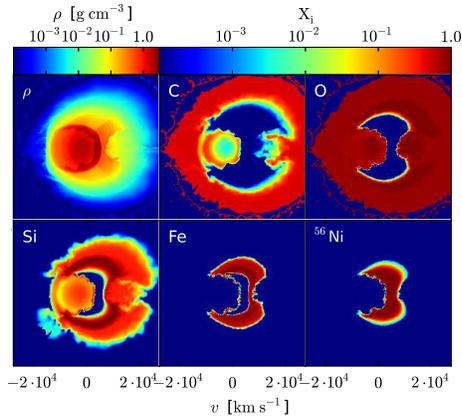}
     \caption{Density and composition of the final ejecta in homologous expansion $100\,\mathrm{s}$ after the explosion
     of a slice in the x-z plane. The mass fraction is shown for carbon, oxygen, silicon, stable iron and radioactive $^{56}\mathrm{Ni}$.}
     \label{fig:comp}
  \end{figure}
  
  The evolution of the binary system is shown in Fig.~\ref{fig:merger}. The mass transfer is stable for more than $15$ orbits.
  After about $600\,\mathrm{s}$ the secondary white dwarf becomes dynamically unstable and is disrupted
  on a timescale of one orbit. As the material of the secondary is accreted violently onto the primary, material
  is compressed and heated up on the surface of the primary white dwarf. As a consequence hotspots form in which carbon
  burning is ignited. When the first hotspot reaches a temperature larger than $2.5 \times 10^9 \mathrm{K}$ at a density of
  about $2 \times 10^6 \mathrm{g\ cm^{-3}}$ we assume that a detonation forms \citep{seitenzahl2009b}. Note
  that despite the high resolution we use, we still tend to underestimate the temperature in the hotspot compared to even
  higher resolution simulations \citep{pakmor2011b}. Only future detailed investigations might be able to decide whether
  or not a detonation really forms but the conditions in our SPH simulations suggest that it is plausible.
  
  At this time we map the whole state of the simulation on a uniform Cartesian grid with a resolution of $768^3$
  grid cells and a box size of $4 \times 10^9\, \mathrm{cm}$. About $0.03 \mathrm{M}_\odot$ of material is lost by
  the mapping as it is outside the box. Since this material makes up less than two percent of the total mass and has a density
  too low to contribute significantly to nuclear burning (i.e. it stays unburned) it does not affect the dynamics
  of the ejecta or the synthetic observables derived from the model.
  
  With these initial conditions we use the \textsc{leafs} code
  \citep{reinecke1999b} that applies the level-set technique to model detonation flames \citep{reinecke1999a,fink2010a}. We
  ignite the detonation at the cell with the highest temperature and follow its propagation through the merged object
  until most of the material is burned. Fig.~\ref{fig:merger} shows that the primary white dwarf and most parts of the secondary
  are already burned $2\, \mathrm{s}$ after the detonation formed. The energy release from nuclear burning
  unbinds the object. Using an expanding grid \citep{roepke2005c} we follow the dynamic ejecta
  until they reach homologous expansion about $100\, \mathrm{s}$ after the detonation was ignited. They have an
  asymptotic kinetic energy of $1.7 \times 10^{51}\, \mathrm{erg}$.
  
  \section{Nucleosynthesis}
  \label{sec:nucleosynthesis}
  
  \begin{figure*}
     \centering
     \includegraphics[width=0.9\linewidth]{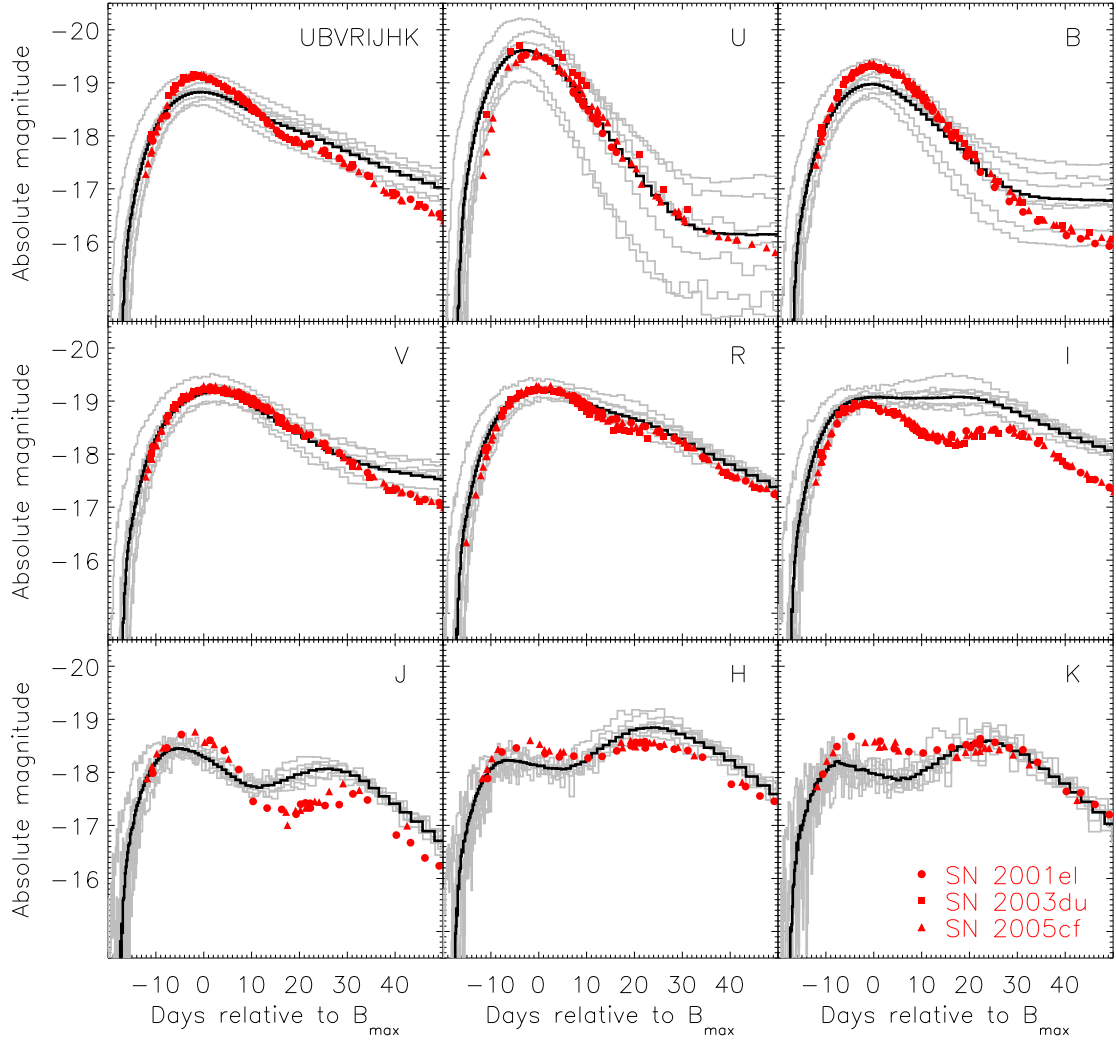}
     \caption{Lightcurves of our model. The panels from top left to bottom right contain UBVRIJHK bolometric and broad-band 
     U,B,V,R,I,J,H,K lightcurves. The black line corresponds to the angle-average of the model. Grey histograms show 
     lightcurves along seven different lines-of-sight representative for the scatter caused by different (100) viewing angles 
     including the most extreme lightcurves. The time is given relative
     to B-band maximum. The red symbols show observational data of three well-observed normal SNe Ia, SN~2001el 
     \citep{krisciunas2003a}, SN~2003du \citep{stanishev2007b}, and SN~2005cf \citep{pastorello2007a}.}
     \label{fig:lightcurves}
  \end{figure*}
  
  \begin{figure}
     \centering
     \includegraphics[width=0.9\linewidth]{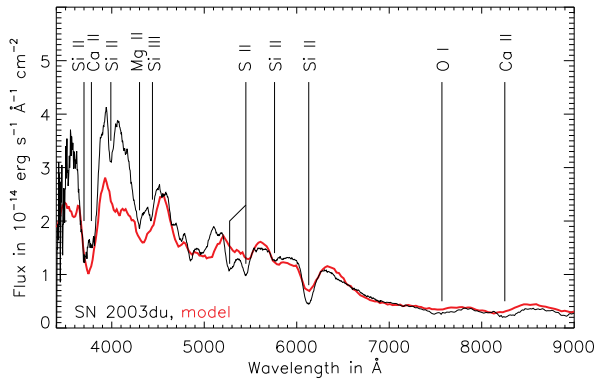}
     \caption{Maximum light spectrum of our model. The red line shows the spectrum of our model one day after maximum
     light in the B-band. The black line shows the observed spectrum of SN~2003du \citep{stanishev2007b} at the same time.}
     \label{fig:spectrum}
  \end{figure}
  
  In order to obtain detailed isotopic abundances of the ejecta we add $10^6$ tracer particles to the simulation
  of the explosion that record their local temperature and density. In a post-processing step we run a 
  detailed nuclear network containing $384$ isotopes on these trajectories \citep{travaglio2004a}. To mimic the
  effect of solar metallicity we choose the initial composition of the tracer particles as $47.5\%\, ^{12}\mathrm{C}$,
  $50\%\, ^{16}\mathrm{O}$ and $2.5\%\, ^{22}\mathrm{Ne}$ by mass.
  
  In the explosion, a total of $0.7\, \mathrm{M}_\odot$ of iron group elements are synthesized. They consist predominantly
  of radioactive $^{56}\, \mathrm{Ni}$ ($0.61\, \mathrm{M}_\odot$) with a small fraction
  of stable $^{58}\, \mathrm{Ni}$ ($0.03\, \mathrm{M}_\odot$) and stable $^{54}\, \mathrm{Fe}$ ($0.02\, \mathrm{M}_\odot$).
  In addition, $0.5\, \mathrm{M}_\odot$ of intermediate mass elements are produced
  in the explosion. The ejecta contain about $0.5 \, \mathrm{M}_\odot$ of oxygen and about
  $0.15\, \mathrm{M}_\odot$ of unburned carbon.
  
  The spatial composition and density structure of the ejecta in homologous expansion are shown in Fig.~\ref{fig:comp}.
  Its inherently three-dimensional structure is a result of the state of the merged object at the time the detonation forms.
  Since the detonation propagates faster at higher densities, the primary white dwarf is burned first and its ashes expand.
  As they expand they sweep around the material of the partially intact secondary white dwarf that is still being burned.
  Therefore, the ashes of the primary have already expanded significantly before burning of the secondary is completed
  (roughly one second later), meaning that the ashes of the secondary completely dominate the center of the ejecta.
  Hence, when the ejecta reach the homologous expansion phase several ten seconds later, the very inner parts of the ejecta
  do not contain material from the primary white dwarf and therefore no iron-group elements.

  \section{Comparison with observations}
  \label{sec:observables}

  \begin{deluxetable}{lcccccc}
    \tabletypesize{\scriptsize}
    \tablecaption{Angle-averaged light curve parameters for selected bands.\label{tab:lc_paras}}
    \tablehead{  & \colhead{Bolometric\tablenotemark{a}} & \colhead{$U$} & \colhead{$B$} & \colhead{$V$} & \colhead{$R$} }

    \startdata
      $t_\mathrm{max}$          & 18.6  & 18.0  & 20.8  & 23.2  & 21.6  \\
      $\Delta m_\mathrm{15}$    & 0.74  & 1.30  & 0.95  & 0.67  & 0.42  \\
      $M(t_\mathrm{max})$       & -19.2 & -19.6 & -19.0 & -19.2 & -19.2 \\
      $M(t_\mathrm{max}(B))$    & -19.1 & -19.5 & -19.0 & -19.2 & -19.2 \\
    \enddata
    
    \tablenotetext{a}{Note that this is not the true bolometric lightcurve of
      our calculations. For comparability to observed bolometric lightcurves, 
      the model lightcurve was reduced to UBVRIJHK bolometric.}
  \end{deluxetable}

  Using the three-dimensional density structure from explosion modeling and the 
  corresponding spatial abundance distribution from the tracer particles we 
  apply the Monte Carlo radiative transfer code \textsc{artis} 
  \citep{kromer2009a,sim2007b} to calculate synthetic spectra and lightcurves 
  for our model.
  
  Fig.~\ref{fig:lightcurves} shows angle-averaged and line-of-sight dependent lightcurves of our model compared with several well-observed normal SNe Ia. The angle-averaged lightcurves show good agreement with observations in the optical bands and reproduce general properties of normal SNe~Ia having a B-band rise time of $20.8$ days, a peak brightness of $-19.6$, $-19.0$, and $-19.2$ in the U, B, and V band, respectively, and a light curve decline rate in the B band of $\Delta m_{15}=0.95$). These values are well within the range of normal SNe~Ia \citep{hicken2009b}. More detailed values for selected bands of the angle-averaged light curves are given in Table~\ref{tab:lc_paras}.
  
  In particular, the angle-averaged lightcurves in the $U$ and $R$ band agree very well with those observed for normal SNe Ia. The angle-averaged synthetic $B$-band lightcurve is about $0.3$ magnitudes fainter at maximum than the three supernovae we compare with and slightly brighter from $25$ days after maximum onwards. The angle-averaged $V$-band lightcurve of our model agrees well with the data up to about $30$ days after maximum, but declines slightly more slowly afterwards.
  
  Due to the asymmetric ejecta structure, our model lightcurves show a non-negligible sensitivity to line-of-sight effects. In the $B$ band, for example, we find peak magnitudes between -19.5 and -18.7. At the same time $\Delta m_{15}(B)$ varies between 0.5 and 1.4, roughly along the Phillips relation \citep{phillips1999a}. For the $U$ band, the scatter is even larger while it decreases in the redder bands. For lower photon energies, the asymmetry of the ejecta is less important, since the optical depths are smaller and photons probe a larger fraction of the total ejecta, thereby making the observables less sensitive to line-of-sight effects (compare \citealt{kromer2009a,kromer2010a}).
  
  The angle-averaged lightcurve of our model in the $I$ band agrees well with the observations up to maximum light. Afterwards it is between $0.2$ and $0.7$ magnitudes brighter than our comparison SNe Ia.
  This offset (in the $I$ band after maximum) can be attributed to a flux excess in the Ca\,{\sc ii} NIR triplet in the synthetic spectra. This could point to an over-abundance of calcium in the model. However, it is more likely to be a shortcoming of our radiative transfer treatment, which uses the simple van Regemorter approximation \citep{vanregemorter1962a} to treat collisional excitation. Thus we likely underestimate the effectiveness of cooling by forbidden transitions of heavy elements and therefore predict that too much cooling occurs via the strong dipole transitions of Ca\,{\sc ii}.

  There is good agreement between model and data in the NIR bands ($J$,$H$,$K$). In particular, our model reproduces time and brightness of the second peaks and even the light curve decline at more than $40$ days after maximum quite accurately. However, high precision modeling of the NIR lightcurves requires an extensive atomic data set to properly simulate flux redistribution by fluorescence \citep{kasen2006b,kromer2009a}. Here, however, we have restricted ourselves for computational reasons to a simplified atomic data set (cd23\_gf-5 of \citealt{kromer2009a}) with only $\sim 400,000$ lines. This has been shown to give reliable results in the optical bands but slightly underpredicts the NIR brightness \citep{kromer2009a} between the primary and secondary peaks.

  Fig.~\ref{fig:spectrum} shows the angle-averaged spectrum of our model one day after B-band maximum. For comparison we overplot an observed spectrum of the normal SN~2003du \citep{stanishev2007b} at the same epoch. 
  Our model shows most of the characteristic features of SNe~Ia, particularly the defining Si\,{\sc ii} doublet at $\lambda\lambda$6347,6371 but also the weaker Si\,{\sc ii} features at $\lambda\lambda$5958,5979 and $\lambda\lambda $4128,4131, the Mg\,{\sc ii} triplet at $\lambda$4481 and the Ca\,{\sc ii} H and K absorption. The S\,{\sc ii} W-feature at $\sim$5400\,\AA\ (which is a blend of several lines) is also visible though it is weaker than in SN~2003du. In the red tail of the spectrum the Ca\,{\sc ii} NIR triplet is clearly visible as well as some indication of the O\,{\sc i} triplet $\lambda\lambda$7772,7774,7775.
  Moreover, the overall spectral shape and the velocity-shifts of most of the line features in the observed spectrum are well reproduced, indicating that the velocity structure of our model is a good representation of that in real SNe Ia.

  \section{Discussion}
  \label{sec:discussion}
  
  The dominant parameter that determines the brightness of the explosion in our model is the mass of the primary
  white dwarf. In the merger the primary  remains mostly unaffected whereas the secondary white dwarf is destroyed. Therefore,
  because the density profile of the primary white dwarf to first order only depends on its mass
  and the density stays high enough to burn to iron-group elements ($^{56}\mathrm{Ni}$) only in the primary white dwarf, the
  brightness of the supernova directly correlates with the mass of the primary white dwarf. 
  
  Physical parameters of secondary importance are the mass of the secondary white dwarf and the composition of both white dwarfs
  (C/O-ratio and metallicity). The material of the secondary is burned only to intermediate-mass elements and oxygen and therefore
  only affects the brightness of the explosion indirectly. The metallicity of the two white dwarfs has only a small impact on the final
  composition of the ejecta by changing the mixture of iron-group elements synthesized \citep{sim2010a}. 
  
  The moment at which the detonation forms is an artificial parameter of our model. Physically it will be determined by the properties of
  the binary system discussed above. Since we cannot resolve the formation of the detonation microscopically, however, we have
  to infer the most likely place and time for it to form from macroscopic properties. Once we assume that a detonation forms on
  the surface of the primary its exact place and time are only secondary effects: the choice for the moment of detonation affects the
  densities at which the material of the secondary white dwarf is burned (since the detonation occurs while the secondary is being destroyed)
  but it does not strongly affect the burning of the primary because it is not dynamically affected by the merger.
  
  Compared to the merger of two $0.9\,\mathrm{M_\odot}$ white dwarfs \citep{pakmor2010a} the merger we present here mainly differs
  in the mass of the primary white dwarf. Since it is considerably more massive, its central density is higher and its material is burned
  predominantly to iron group elements. In addition, the mass ratio of the binary system we model here is well below unity. Therefore, the more
  compact primary is not subject to noticeable tidal forces from the secondary.
  
  In addition, although exact initial conditions may weaken the merger, we show that it still leads to the formation of hot spots on the surface of the
  primary, in which densities and temperatures are sufficiently high that a detonation is plausible. This is consistent with the properties of hot
  spots seen in previous studies with less relaxed initial conditions \citep{pakmor2010a,pakmor2011b} and also with very recent  simulations by 
  \citet{raskin2011a} who found similar hot spots in calculations with relaxed initial conditions and only slightly smaller resolution than our simulation.
  Note that \citet{dan2012a} do not find hotspots in similar systems with exact initial conditions but more than a factor of ten fewer particles.
  
  Our model can be interpreted as a pure detonation of an isolated sub-Chandrasekhar mass white dwarf as described
  in \citet{sim2010a} with an additional component from the remains of the secondary white dwarf surrounding the system. Quantitatively,
  however, the color lightcurves of our model agree significantly better with observations than the toy models of \citet{sim2010a}.
  In particular, our model also shows good agreement in the NIR-bands and reproduces the position of the secondary
  peaks in these bands very well. 
  
  This difference to the toy models is likely to be associated with a different density profile and composition structure in the inner parts
  of the ejecta. As discussed in Section \ref{sec:nucleosynthesis} this is a result of the presence of the secondary white dwarf
  and its interaction with ashes of the primary when they expand. Also, the asymptotic kinetic energy per mass of the ejecta is smaller
  for our merger model than for detonations of isolated sub-Chandrasekhar mass white dwarfs, because the burning of the
  secondary white dwarf is less complete and hence the average energy release per mass is smaller.
  
  Overall, our explosion reproduces observational data of normal SNe Ia of the same brightness reasonably well. Since the time span from the
  onset of mass transfer between the two white dwarfs until the ejecta reach homologous expansion is very short (only of the order 
  of a few minutes) there is no extended outer envelope the explosion ejecta interact with later than a few minutes after the explosion, in 
  contrast to the model by \citet{fryer2010a}.
  
  Observational constraints \citep{napiwotzki2005a,napiwotzki2007a} on the number of double white dwarf binaries significantly more massive
  than a Chandrasekhar mass are still not very restrictive.
  If violent mergers contribute noticeably to the total SNe Ia rate, our calculations suggest that it will be hard to 
  distinguish the merger events from the bulk of normal SNe Ia via optical/NIR lightcurves and optical spectra. Further detailed
  work is therefore needed to investigate whether there are other characteristic properties of SN Ia explosions from violent white dwarf
  mergers and to explore how the observational display changes with the properties of their progenitor system.  
  
  \section*{Acknowledgements}
    This work was supported by the Deutsche Forschungs\-gemeinschaft via
    the Transregional Collaborative Research Center TRR 33 ``The Dark
    Universe'', the Excellence Cluster EXC153 ``Origin and Structure
    of the Universe'' and the Emmy Noether Program (RO 3676/1-1). Parts of the simulations
    were carried out at the John von Neumann Institute for Computing (NIC) in J\"{u}lich, Germany
    (project hmu14). R.~P. gratefully acknowledges financial support of the Klaus Tschira Foundation.

  \bibliographystyle{aa}

\end{document}